\documentstyle[editedvolume,numreferences,epsf]{crckapb} 
\epsfverbosetrue

\begin{opening}
\title{THE (UNSTABLE) THRESHOLD OF BLACK HOLE FORMATION}

\author{M.W. CHOPTUIK}
\institute{Center for Relativity\\
           Dept. of Physics \\
           The University of Texas at Austin\\
           Austin TX, 78712-1081}

\end{opening}

\begin{document}


\section{Introduction\label{SEC_INTRO}}
In recent years it has become apparent that intriguing phenomenology
exists at the threshold of black hole formation in a large class of
general relativistic collapse models.  This phenomenology, which includes
scaling, self-similarity and universality, is largely analogous to
statistical mechanical critical behaviour, a fact which was first noted
empirically, and subsequently clarified by perturbative
calculations which borrow on ideas and techniques from dynamical
systems theory and renormalization group theory.  This contribution,
which closely parallels my talk at the conference,  consists
of an overview of the considerable ``zoo''' of critical solutions
which have been discovered thus far, along with a brief discussion
of how we currently understand the nature of these solutions 
from the point of view of perturbation theory.  The reader who
wishes additional details concerning the subject
is referred to Gundlach's excellent recent 
review~\cite{GUN_REVIEW}.
Earlier synopses by Evans~\cite{EVA93} and Eardley~\cite{EAR93} may 
also be of interest.  Finally, those readers interested in 
the relationship of black-hole critical phenomena to questions
of cosmic censorship---a topic which is {\em not} discussed 
below---can consult Wald's recent discussion of the 
status of cosmic censorship~\cite{WAL97}.

\section{Critical Behaviour in Massless Scalar Collapse\label{SEC_EMKG}}

I begin with a fairly detailed description of critical 
behaviour in the self-gravitating dynamics of a spherically-symmetric, 
massless scalar field.
Apart from historical reasons, I do so, not only because it is arguably 
the simplest ``realistic''
model of gravitational collapse, but, more importantly, 
because most of the features we currently associate with critical collapse 
can be seen in the model.

\subsection{Equations of Motion\label{SEC_EMKG_EOM}}
Adopting the usual spherical coordinates $(t, r, \theta, \varphi)$,
and geometric units, $G=c=1$, the spherically-symmetric
spacetime metric can be written
\begin{equation}
	ds^2 = -\alpha^2(r,t) \, dt^2 + a^2(r,t) \, dr^2 + r^2 \left( d\theta^2 + 
                                    \sin^2\theta \, d\varphi^2 \right)
  \, .
\label{FLATMETRIC}
\end{equation}
The coordinate system I have thus adopted is a natural generalization 
of Schwarzschild coordinates---in numerical relativity parlance 
it is the polar-areal (or polar-radial) system.  Note that the 
radial coordinate, $r$, measures proper surface area and thus 
has an immediate geometric interpretation.  The time coordinate, $t$,
on the other hand, has no particular physical significance.  However,
from the point of view of critical phenomena, there is a preferred 
labeling (reparametrization) of the $t={\rm constant}$ surfaces, namely
the one given by the proper time, $T_0(t)$, of an observer at rest 
at $r=0$:
\begin{equation}
	T_0(t) \equiv \int_0^t \alpha(0,{\tilde t}) \, d{\tilde t}  \, .
\label{DEFT0}
\end{equation}
Defining auxiliary scalar field variables, $\Phi$ and $\Pi$:
\begin{eqnarray}
	\Phi(r,t)  & \equiv & \frac{\partial \phi}{\partial r}(r,t)  \, ,
\label{DEFPHI} \\
	\Pi(r,t)   & \equiv & \frac{a}{\alpha} 
                       \frac{\partial \phi}{\partial t}(r,t)  \, ,
\label{DEFPI}
\end{eqnarray}
a sufficient set of equations for the EMKG (Einstein-massless-Klein-Gordon) 
model is 
\begin{eqnarray}
	\frac{\partial \Phi}{\partial t} & = &
		\frac{\partial}{\partial r} \left( \frac{\alpha}{a} \Pi \right)
   \, ,
\label{PHIDOT} \\
	\frac{\partial \Pi}{\partial t} & = &
		\frac{1}{r^2}
		\frac{\partial}{\partial r} \left( r^2 \frac{\alpha}{a} \Phi \right)
   \, ,
\label{PIDOT} \\
	\frac{1}{\alpha} \frac{d\alpha}{dr} & - & \frac{1}{a}\frac{da}{dr} +
	\frac{1-a^2}{r} = 0
   \, ,
\label{SLICE} \\
	\frac{1}{a} \frac{da}{dr} & + & \frac{a^2-1}{2r} -
		2\pi r\left( \Pi^2 + \Phi^2 \right) = 0
   \, .
\label{HAMC}
\end{eqnarray}
Here~(\ref{SLICE}) is the {\em slicing condition} which constrains 
the lapse function, $\alpha$, at all instants of time, and~(\ref{HAMC})
is the {\em Hamiltonian constraint} which similarly constrains the 
radial metric function $a$.

It is useful to discuss the dynamics of the scalar field in terms 
of further auxiliary variables, $X$ and $Y$, defined by
\begin{eqnarray}
	X(r,t) & \equiv & \sqrt{2\pi} \frac{r}{a} \Phi =
          \sqrt{2\pi} \frac{r}{a} \frac{\partial\phi}{\partial r} \, ,
\label{DEFX} \\
	Y(r,t) & \equiv & \sqrt{2\pi} \frac{r}{a} \Pi =
          \sqrt{2\pi} \frac{r}{\alpha} \frac{\partial\phi}{\partial t} \, .
\label{DEFY}
\end{eqnarray}
Note that the equations of motion~(\ref{PHIDOT})--(\ref{HAMC}) are 
invariant under the trivial rescalings $r\to k r$, $t \to k t$, for arbitrary 
$k > 0$, and that $X$ and $Y$ are form-invariant under such 
transformations.  Is is also convenient to 
introduce the {\em mass aspect}
function, $m(r,t)$, defined in analogy with the usual Schwarzschild form of 
the static spherically-symmetric metric:
\begin{equation}
	a^2(r,t) = \left( 1 - \frac{2 m(r,t)}{r} \right)^{-1} \, .
\label{DEFM}
\end{equation}
In terms of the $X$ and $Y$ variables, $dm/dr = X^2 + Y^2$,
and the total mass (ADM mass), $M_{\rm ADM}$, of the spacetime is given
by
\begin{equation}
	M_{\rm ADM} = \int_0^\infty  \frac{dm}{dr} \, dr = 
                 \int_0^\infty \, X^2 + Y^2 \, dr \, .
\label{ADMMASS}
\end{equation}
Another useful relationship expresses the spacetime curvature
scalar, $R$, as a function of $X$ and $Y$:
\begin{equation}
	R = - 8 \pi T = 8 \pi \nabla^\mu \phi \nabla_\mu \phi =
     \frac{4}{r^2} \left( X^2 - Y^2 \right) \, .
\label{R}
\end{equation}
where $T$ is the trace of the scalar field stress-energy tensor.

\begin{figure}
\begin{center}
		\epsfxsize=120mm
      \epsfbox{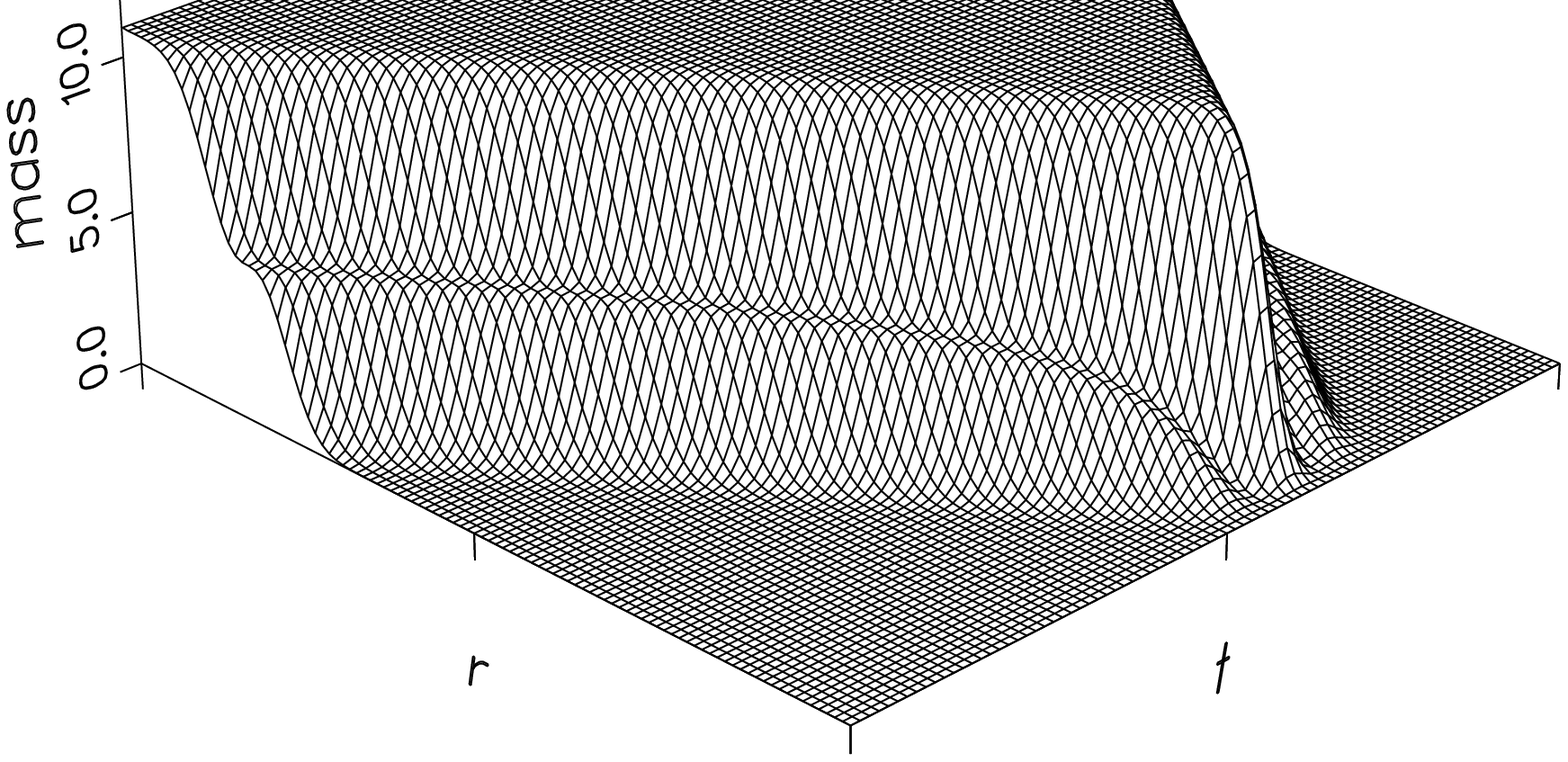}
		\epsfxsize=120mm
      \epsfbox{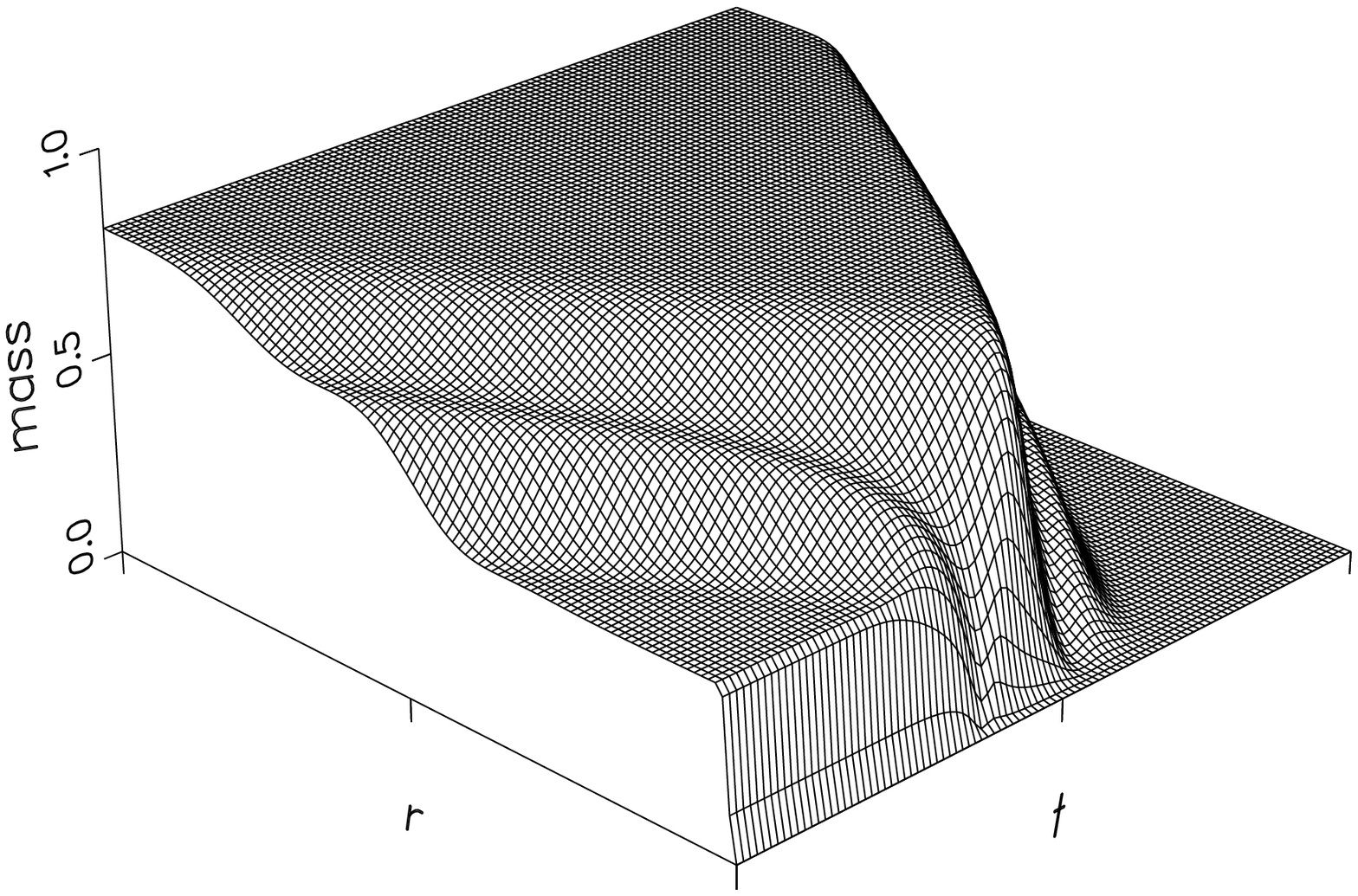}
\end{center}
\vspace{-170mm}
\caption{Behaviour of the mass aspect function, $m(r,t)$, for cases 
of complete dispersal (top), and black hole formation (bottom), in
massless scalar collapse.
The origin, $(0,0)$, in both plots is at the extreme right.
Initial data in both cases is a single ingoing 
Gaussian profile---$\phi(r,0) = p \, g(r)$---the calculations differ
only in the choice of the overall amplitude factor, $p$, which 
is sub-critical ($p < p^\star$) in the first instance, and 
super-critical ($p > p^\star$) in the second.  Note that flat 
regions in the plots are vacuum.  In the sub-critical case,
the scalar field implodes through the origin, then completely 
disperses leaving (essentially) flat spacetime in its wake. 
In the super-critical case, the scalar field again implodes 
through the origin, but now forms a black hole containing 
roughly half of the total mass of the spacetime. Dynamically, the 
geometry within the dispersing (outgoing) scalar field settles 
down to an exterior-Schwarzschild solution on a time scale 
set by the size of the hole, $T \sim  R_{\rm BH} = 2M_{\rm BH}$.
}
\label{FIG1}
\end{figure}

As is well known, polar-areal coordinates cannot cross apparent
horizons, and thus, for the most part, cannot penetrate 
event horizons.   However, black hole formation is clearly 
signaled in a calculation~(see Figure~\ref{FIG1}) by 
(among other things)
\begin{equation}
	\left. \frac{2m}{r} \right\vert_{R_{\rm BH}} \to 1 \, ,
\end{equation}
at some radius $r=R_{\rm BH}$ from which the mass, 
$M_{\rm BH} = R_{\rm BH}/2$, of the final black hole can be quite
accurately estimated.  

\begin{figure}[t]
\epsfxsize=100mm
\epsfbox{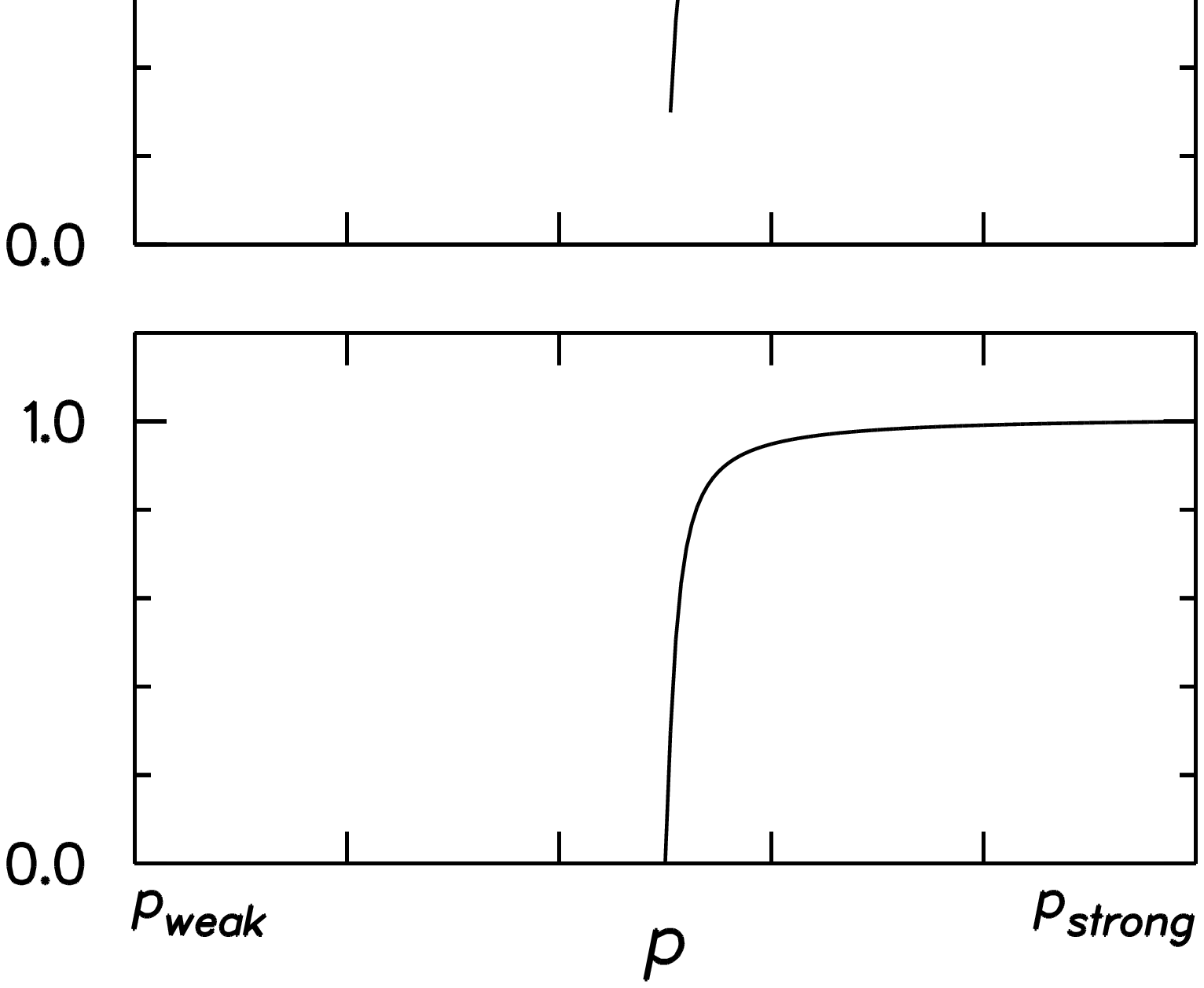}
\vspace{-30mm}
\caption{Schematic illustration of black-hole-threshold possibilities. The 
top panel represents a first-order (Type I) transition, where 
the ``order parameter'' (the black hole mass) exhibits a gap at 
threshold, while the bottom panel shows a second-order (Type II) transition,
where the black hole mass is infinitesimal at the critical point.}
\label{FIG2}
\end{figure}

\subsection{Competition and the Threshold of BH Formation
\label{SEC_EMKG_COMP}}
At a heuristic level, the existence of critical behaviour
in the EMKG system and other models is a direct result of
{\em competition} in the dynamics.  The nature of this 
competition can be seen by addressing the question:
``given generic initial data representing an 
imploding pulse (shell) of scalar radiation, where can the 
energy in the system end up at late times?'' Roughly 
speaking, the kinetic energy of the massless field wants 
to disperse the field to infinity, whereas the gravitational
potential energy (entirely self-induced), if sufficiently 
dominant during the collapse, will result in the trapping of some amount 
of the mass-energy of the system in a black hole.  Indeed, the fact
that, for {\em generic} initial data,  there {\em are} only these 
two qualitatively distinct end-states in the model has 
been rigorously established by Christodoulou~\cite{CHR,WAL97}.  However,
empirical evidence (i.e. direct solution of the equations of 
motion) rapidly leads one to the same conclusion.   The key
point is that the dynamical competition can be controlled by
tuning a parameter in the initial conditions: it is an easy matter 
to set up families of initial data, $\{ \Phi(r,0;p), \Pi(r,0;p) \}$
such that if the parameter, $p$, is less than some 
critical (threshold) value, $p^\star$, the scalar field completely 
disperses, while if $p > p^\star$ a black hole forms 
(Figure~\ref{FIG1}).

\begin{figure}
\begin{center}
\epsfxsize=130mm
\epsfbox{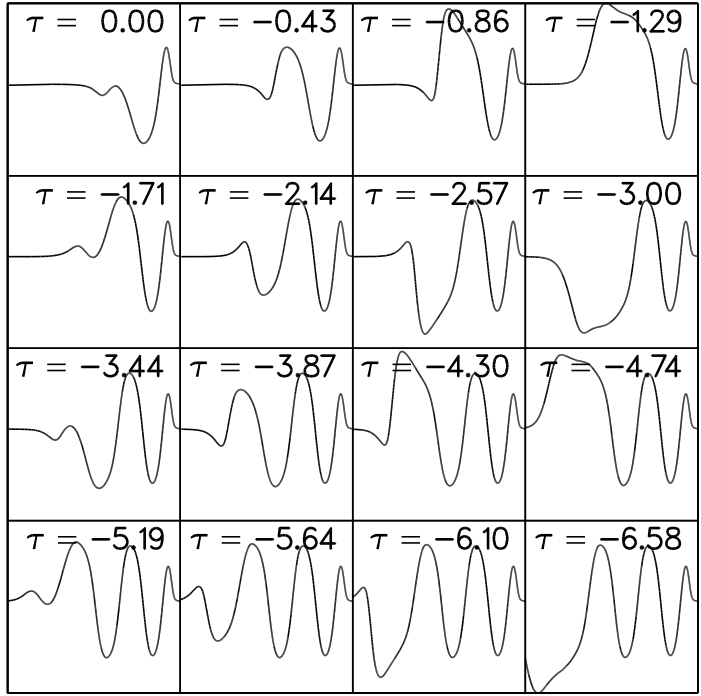}
\vspace{-75mm}
\end{center}
\caption{Near-critical evolution of the auxiliary scalar 
field variable, $X$ (see equation~(\ref{DEFX})), 
in massless scalar collapse. 
Evolution proceeds left-to-right, top-to-bottom.  In each
frame, $X$ is plotted versus the logarithmic radial coordinate,
$\rho \equiv \ln r$, with displayed ordinate values ranging
from -0.37 to 0.33.  Constant increments of logarithmic time,
$\tau \equiv \ln (T_0^\star - T_0)$, separate each successive
snapshot.  The discrete self-similarity of the critical solution
can be clearly seen by comparing the trailing (left-most) edges 
of pairs of waveforms separated by two rows in the plot. Each
such pair is very nearly in the same ``phase'' of critical 
evolution, but the dynamics of the later frame (more negative
$\tau$) is occurring on a scale some $\exp(\Delta) \approx 30$ times 
smaller than that of the earlier snapshot.  Note that the 
leading parts of the waveforms (the roughly sinusoidal oscillations)
are almost purely-outgoing, and do not, {\em per se}, constitute 
part of the critical dynamics. The fact that these oscillations
appear to be ``frozen'', rather than propagating, is a result of the 
fact that the time between successive frames is exponentially decreasing.
At any scale (position along the horizontal axis) of the critical evolution,
there are three basic possibilities:  (1) if the evolution is sub-critical
(at that scale), the scalar field will completely disperse, 
(2) if the evolution is super-critical, a black hole 
{\em with a size set by the current scale of the critical dynamics}  
will form, and (3) if the evolution is 
precisely-critical, the strong field evolution will continue to 
a smaller scale.  This behaviour provides convincing evidence for 
a Type II transition in the model.  Plots of the other scalar
field variable, $Y$, as well as $X^2 + Y^2 = dm/dr$ and 
$X^2 - Y^2 = 4 R /r^2$ show analogous oscillatory behaviour between
fixed limits.  Since the strong-field regime is characterized by $r\to 0$ for 
a precisely critical solution, this last fact shows immediately that the
spacetime scalar curvature, $R$, diverges in the critical limit.
}
\label{FIG3}
\end{figure}

The existence of interpolating families and black hole thresholds
in the EMKG model was well-established in early studies 
of the system~\cite{GOL87,CHO86,CHO89}.
The discovery of critical behaviour in the model,
however, came later, and was a direct result of a question
Christodoulou posed in 1987~\cite{CHR_PC}: ``will black hole 
formation turn on at {\em finite} or {\em infinitesimal} mass 
for a generic interpolating family at threshold?'' (see Figure \ref{FIG2}).
Viewing the black-hole mass as an order parameter, I refer to these 
two possibilities as Type I and Type II transitions, respectively,
in analogy with first and second order phase transitions 
in statistical mechanical systems.

\subsection{Type II Behaviour in the EMKG Model
\label{SEC_EMKG_TYPEII}}
Detailed phenomenological studies~\cite{CHO92,CHO93,CHO94}
using many initial data 
families clearly demonstrated that the strong-field (i.e. critical) dynamics
in the EMKG model is characterized by an essentially {\em unique}
solution of the equations of motion, which I denote schematically
as $Z^\star$.  Like the Schwarzschild solution, $Z^\star$, is 
only determined upto an overall (length/time) scale, but unlike
Schwarzschild, the critical solution is obviously (by construction!)
{\em unstable}---the slightest perturbation will result in 
either complete dispersal or black hole formation.  

A key feature of the critical solution (and of all currently known 
Type II solutions) is {\em self-similarity} or {\em scale 
invariance}---loosely speaking, the critical solution 
has a scale (homothetic) Killing vector. 
(Correspondingly, the 
known Type I critical solutions discussed in section~\ref{SEC_EYM}
are characterized by timelike 
Killing vectors).  Empirically, this means as one tunes closer and closer 
to a critical point one sees, in a single evolution, the same strong-field
dynamics playing out on a wider and wider range of spatial scales, 
always in a (shrinking) neighborhood of $r=0$.  In the scalar field 
case (see Figure~\ref{FIG3}), the dynamics at any particular ``scale epoch''
is decidely non-trivial---i.e. the self-similarity is {\em discrete},
rather than continuous.  This novel ``echoing'' feature
of the solution, whereby the dynamics repeats on scales related 
by a factor of $e^\Delta$, $\Delta=3.44...$, is one of the most 
intriguing aspects of the solution, and its origin 
remains somewhat of a mystery.  We can express
the discrete self-similarity is slightly more mathematical 
terms by noting that in 
a precisely critical evolution, the strong-field evolution
``accumulates'' at a singular event with $(r,T_0)$ 
coordinates $(0,T_0^\star)$.  Then defining logarithmic
coordinates, $\rho\equiv\ln r$, $\tau\equiv\ln(T_0^\star -T_0)$, 
we have
\begin{equation}
	Z^\star\left( \rho \pm n \Delta, \tau \pm n \Delta \right) 
   \sim Z^\star \left( \rho,\tau \right) 
  \quad\quad n = 0, 1, 2, \, \cdots \, .
\end{equation}

\begin{figure}[t]
\begin{center}
\epsfxsize=120mm
\epsfbox{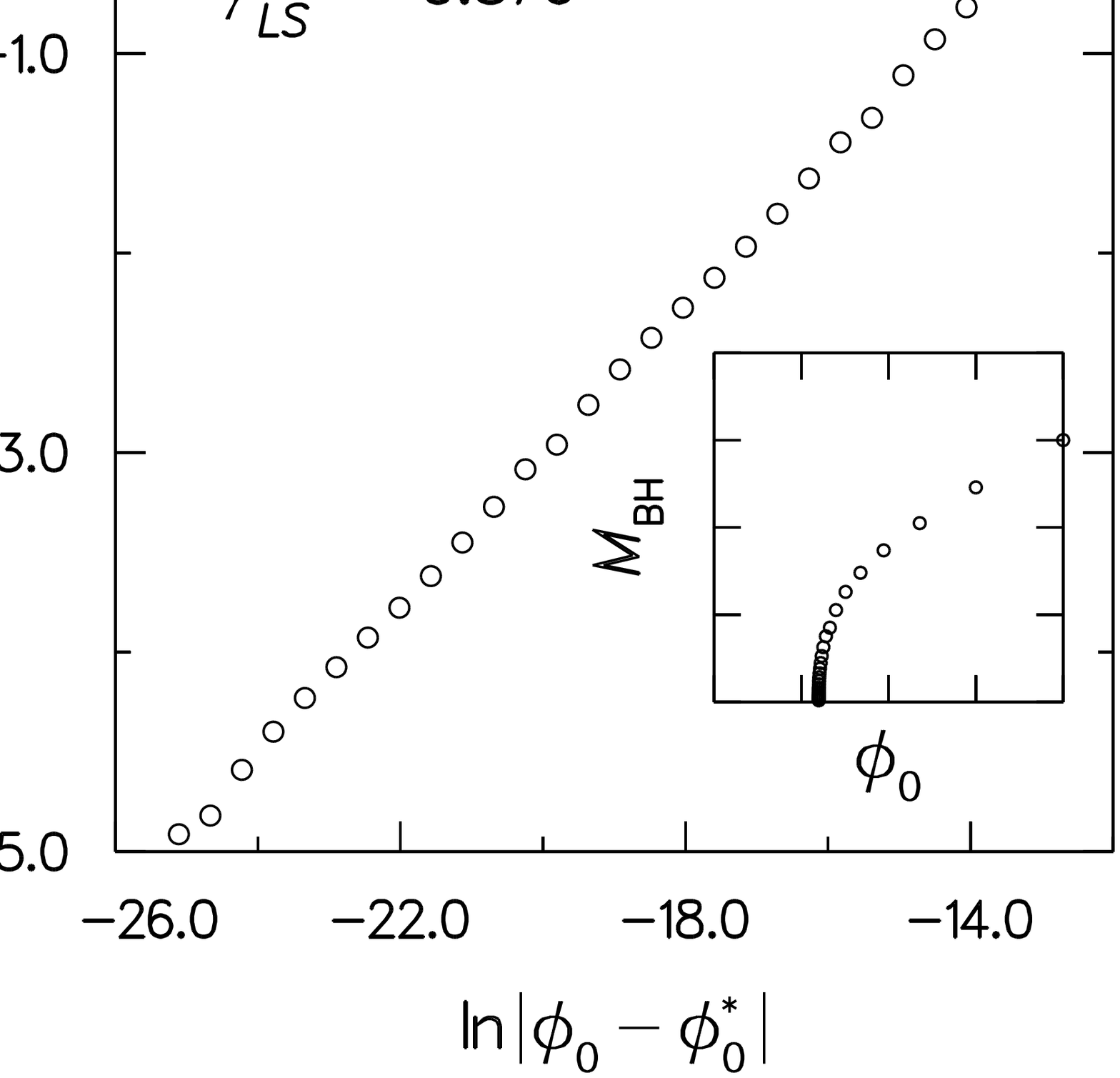}
\vspace{-40mm}
\end{center}
\caption{Typical evidence for mass scaling in the collapse of 
a massless scalar field.  The data displayed here was generated 
from a one-parameter family of initially-ingoing Gaussian pulses 
of scalar field in which the overall amplitude of the pulse was 
varied. The mass-scaling exponent, $\gamma_{\rm LS} \approx 0.376$,
was determined from a least-squares fit.  The inset clearly shows 
the ``second order'' (Type II) nature of the transition.  It is 
worth noting that, particularly for compact pulse shapes such 
as the Gaussians used here, the mass scaling persists well out 
of the asymptotic regime, $p \to p^\star$. 
}
\label{FIG4}
\end{figure}

As explained in more detail in Figure~\ref{FIG3}, the nature of the critical 
dynamics (which is now well-understood, albeit in a non-rigorous
fashion, in perturbation theory---see section~\ref{SEC_PERT}), 
makes it clear that tuning 
of initial data provides the mechanism for making arbitrarily 
small black holes, which in itself establishes that Type 
II transitions occur generically in the EMKG model.  Direct 
measurement of the black hole masses in the supercritical regime
provides additional evidence.  One finds that, for a generic interpolating
family $S[p]$, and as $p\to p^\star$, the black hole 
masses are well-fit by a scaling law~(see Figure~\ref{FIG4}):
\begin{equation}
	M_{\rm BH} = c_f \vert p - p^\star \vert^\gamma \, ,
\label{MASSSCALE}
\end{equation}
where $c_f$ is a family-dependent constant, but $\gamma = 0.37...$ is 
universal (i.e. family-independent).

As also noted in Figure~\ref{FIG3}, the nature of the critical solution 
immediately implies that the scalar curvature (see equation~(\ref{R}))
grows without bound near $r=0$ in a precisely critical evolution. 
It is also clear from the simulations that these regions of 
arbitrarily high curvature are visible by observers at infinity.
Indeed, the trailing edge of the scalar field pulse itself effectively 
does the job of tracking outwards propagating null geodesics emitted 
from the critical region.
Again, the point is that, contrary to some expectations~\cite{HAM96a}, 
without the formation of a horizon, one cannot indefinitely localize
the mass-energy of the scalar field.  Finally, I note that all
of the above features of critical spherically-symmetric scalar collapse 
have been reproduced---using a variety of theoretical approaches 
and numerical techniques---by several other 
groups~\cite{HAM96a,GAR95,HOD_FS,HOD_CSF}.

\section{Other Early Results in Critical Collapse\label{SEC_EARLY}}
\subsection{GRAVITATIONAL WAVE COLLAPSE\label{SEC_EARLY_GW}}
Shortly after the discovery of critical behaviour in the EMKG model,
Abrahams and Evans~\cite{ABR93} presented exciting results demonstrating that 
very similar effects occurred in the purely gravitational (i.e. 
vacuum) collapse of {\em axisymmetric} gravitational waves.  Again, at
least in  principle, it is straightforward to construct interpolating
families in this case, but, computationally, this problem is 
much more difficult to treat than the spherically symmetric EMKG system
described above.  Nonetheless, Abrahams and Evans were able to 
produce convincing evidence for (1) 
a discretely self-similar threshold solution, this time with 
$\Delta\approx 0.6$, and (2) mass-scaling in the super-critical
regime, with $\gamma \approx 0.37$.  The work in~\cite{ABR93} used 
a single interpolating family; a follow-up paper~\cite{ABR94} reported 
the observation of similar results from a second family, thus 
providing evidence for the universality of their critical solution.

\subsection{RADIATION FLUID COLLAPSE\label{SEC_EARLY_FLUID}}
A third and very important example of critical behaviour came 
with the work by Evans and Coleman~\cite{EVA94}, who studied 
the spherically symmetric collapse of a perfect fluid with
the simple equation of state
\begin{equation}
	P = \frac{1}{3} \rho \, ,
\end{equation}
where $P$ and $\rho$ are the fluid's pressure and density
respectively.  Early on, Evans had realized that the self-similar
nature of the Type II transitions which were being observed
could be used to great advantage in understanding what was 
happening in critical collapse.  
Familiar with the 
existence of continuously self-similar solutions in relativistic
fluid flow, he argued that that the phase transition in the 
case of radiation fluid collapse should be characterized 
by continuous self-similarity.  Proceeding from the {\em ansatz}
of self-similarity, Evans and Coleman constructed a 
precisely self-similar solution and verified that it was 
identical to the critical solution generated from full
dynamical evolution (again using interpolating families).
Once more, convincing evidence for mass-scaling in the 
super-critical regime, with $\gamma \approx 0.36$, was 
found.  Importantly, Evans also suggested that an
investigation of the perturbative mode structure of the 
critical solution could provide the basis for the 
computation of the mass-scaling exponent.  Finally, the 
apparent numerical equality of $\gamma$ for the EMKG,
vacuum axisymmetric and radiation fluid cases, led to 
a provocative, but short-lived conjecture, that there 
might be ``true'' universality of $\gamma$ across various
collapse models~\cite{EVA94}.

\section{Self Similarity in Critical Collapse\label{SEC_SS}}
As Evans anticipated, the self-similar nature of 
Type II critical solutions {\em has} proven crucial to 
our current understanding of black hole critical phenomena. 
Here I will only briefly sketch some key ideas concerning 
self-similarity in this context---the interested reader can consult Gundlach's 
review~\cite{GUN_REVIEW} (and references contained therein) for a much
more thorough presentation. 

\begin{figure}
\centerline{\epsfxsize=100mm \epsfbox{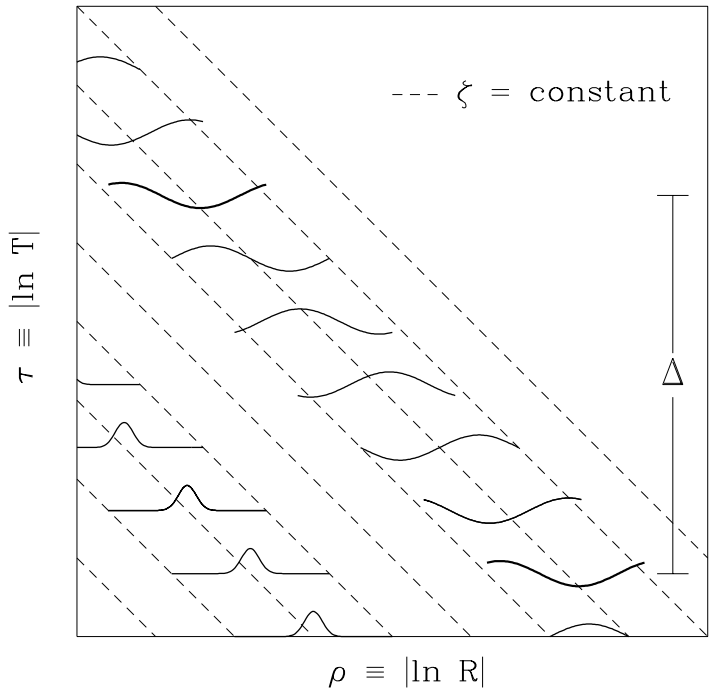}}
\caption{Schematic illustration of continuous (CSS) and 
discrete (DSS) self-similarity in critical collapse. 
The similarity variable $\zeta \equiv R / T$ is constant 
along dashed lines.  In the case of CSS evolution (left 
part of the plot), a {\em single} critical profile propagates
to the left (i.e. to smaller spatio-temporal scales), while in the 
DSS case, a non-trivial, but periodic, {\em sequence} of profiles, 
is generated.  The periodicity in log-space, $\Delta$, is a
model-dependent (but initial-data-independent) scaling exponent.
}
\label{FIG5}
\end{figure}

Restricting attention to spherical symmetry, we
choose ``geometric'' coordinates, $R$ and $T$,  which are
adapted to any particular critical solution.  In the 
context of polar-radial gauge discussed above, obvious choices 
are $R\equiv r$ (areal radius) and $T \equiv ( T_0^\star - T_0 )$,
so that the $(0,0)$ is the central singularity of the precisely 
critical solution, and time advances to the {\em past} of the 
singularity.  The natural similarity variable (coordinate) for a type II 
critical solution is then
\begin{equation}
\zeta = \frac{R}{T},
\end{equation}
and one can use either $T$ or $R$, or more conveniently,
$\tau\equiv\ln T$ or $\rho\equiv\ln R$ as a second coordinate.  Evolution
along lines of constant $\zeta$ (see Figure~\ref{FIG5}) then represents 
evolution in scale.  Continuously self-similar (CSS) solutions
are characterized by 
\begin{equation}
	g\left(\zeta,\tau\right) = g\left(\zeta,\tau'\right)
	\quad\quad \tau \, , \, \tau' \,\, {\rm arbitrary},
\end{equation}
where $g$ denotes any dynamical variable which exhibits scaling.
Discretely self-similar (DSS) solutions, on the other hand, satisfy
\begin{equation}
	g\left(\zeta,\tau\right) = g \left( \zeta,\tau \pm n \Delta\right)
	\quad\quad n = 0, 1, 2, \, \cdots \, ,
\end{equation}
where $\Delta$ is the model-dependent ``echoing exponent''.
It is important to note that, in both cases, the self-similar 
solution is {\em not} a complete spacetime, but (in $R$, $T$
coordinates) a wedge-shaped region 
which, one expects, can be analytically continued
(see~\cite{GUN_EMKG,HIR95a}) to produce a complete spacetime.

\section{Perturbation Theory: Mode Structure of Critical Solutions
\label{SEC_PERT}}
Again, as Evans suggested~\cite{EVA94}, the application of perturbation theory 
to black-hole threshold solutions has proven to be a powerful 
technique for studying and explaining many of the key features 
of critical collapse.  Briefly, for any given collapse model
(which includes specification of the matter content,  coupling
to gravity and other fields, symmetries, etc.), all of the 
current evidence suggests that one will find one or more 
isolated (in ``initial-data space'') critical solutions, $g^\star$.
Operationally, $g^\star$ can be constructed either (1) indirectly, using
the partial differential equations of motion and interpolating
families of initial data, or (2) directly, starting from 
an {\em ansatz} which reflects the particular type of self-similarity 
exhibited by the critical solution.

Schematically (see \cite{GUN_REVIEW} and \cite{KOI96}
for more details), one then considers 
perturbations about the background, $g^\star$, and looks for 
eigenmodes, $f_i(\zeta)$ with corresponding eigenvalues, $\lambda_i$,
with the assumption that (small) departures, $\delta g(\zeta,\tau)$, from the 
critical solution can be then be expressed as
\begin{equation}
	\delta g(\zeta,\tau) \equiv g(\zeta,\tau) - g^\star(\zeta,\tau)
   = \sum_i C_i \exp \, (\lambda_i \tau) \, f_i ( \zeta ) \, ,
\label{MODES}
\end{equation}
where the $C_i$ are coefficients.  

In terms of explaining the mass-scaling and universality of the 
Type II transitions discussed above, the crucial observation
was made by Koike, Hara and Adachi~\cite{KOI95}, who noted 
that the ``sharp'' nature of the transition (as well as the 
universality) strongly suggested that of all the modes, only 
{\em one}, $f_\star$, has an associated eigenvalue, $\lambda_\star$
with
\begin{equation}
	{\rm Re} \, \lambda_\star > 0 .
\end{equation}
That is, {\em there is only one growing mode associated with 
the critical solution}---all other modes generically die off as the 
critical solution is approached.  It is then a straightforward 
matter of linearization of~(\ref{MODES}) and dimensional 
analysis to show that the black hole mass will satisfy a 
scaling law
\begin{equation}
	M_{\rm BH} \propto 
              \left\vert p - p^\star \right\vert^{1/{\rm Re}\,\lambda_\star}
   \, ,
\end{equation}
so that 
\begin{equation}
	\gamma = \frac{1}{{\rm Re}\,\lambda_\star} \, . 
\end{equation}
This argument provides an immediate and intuitive explanation
for the observed universality (initial-data independence) of 
the Type II critical solutions---in tuning the family parameter,
$p$, to the threshold value $p^\star$, one is effectively tuning 
out the unstable mode in the initial conditions, so that 
as $p\to p^\star$, the strong-field dynamics is well-described 
by the precisely critical solution over more and more 
decades of scale. Moreover, the eventual departure from 
$g^\star$ in any specific computation (sub-critical or 
super-critical) is well-approximated by 
\begin{equation}
	\delta g(\zeta,\tau) \sim \exp(\lambda_\star \tau) \, f_\star(\zeta) \, ,
\end{equation}
{\em independently} of any specifics of the initial data.  Thus,
although Type II critical solutions are clearly unstable, they
are, in some sense, the {\em least} unstable solutions 
possible, by virtue of the fact that they possess only a
single growing mode.

Koike~{\em et al} tested their ideas in the context of 
the radiation fluid critical solution, found 
strong evidence that there {\em was} only one growing 
mode in perturbation theory, and from the eigenvalue of 
the mode, computed a scaling exponent, $\gamma = 0.3558\ldots$ in 
excellent agreement with the Evans and Coleman simulations.
Concurrent with~\cite{KOI95}, Maison~\cite{MAI95} presented 
similar calculations which were, moreover, generalized to the 
case
\begin{equation}
	P = k \rho   \quad\quad 0.01 \le k \le 0.888  \, .
\end{equation}
His specific results for $k=1/3$ were in precise agreement with Koike 
{\em et al}, and, moreover, his general results clearly showed that the 
mass-scaling exponent was dependent on $k$---ranging from 
$\gamma = 0.114$ for $k=0.01$ to $\gamma = 0.852$ for $k=0.888$. 
This, of course, provided strong evidence for the model-{\em dependence} (i.e.
``non-universality'') of $\gamma$.

The picture of Type II critical solutions as ``intermediate
attractors''~\cite{BAR79} with 1-dimensional unstable 
manifolds has now been validated for many spherically 
symmetric models, including axion/dilaton 
collapse~\cite{EAR95,HAM96b} (see also~\cite{LIE96} for 
dynamical evolutions in a generalized setting),  
the EMKG model~\cite{GUN_EMKG,GUN95},
the SU(2) Einstein-Yang-Mills model briefly discussed below~\cite{GUN_EYM},
and scalar electrodynamics~\cite{GUN_CHARGE}.   In the last 
case, Gundlach and Martin-Garcia also predicted an additional 
scaling law for the (scalar) charge of black holes, which was 
very quickly observed in independent work by
Hod and Piran~\cite{HOD_CSF}.

In addition to the critical solutions with one unstable mode
described above, other self-similar, strong field 
solutions have been constructed and investigated (see 
for example, the work by Hirschmann and Eardley~\cite{HIR95a,HIR95b}).
It seems reasonable to assume that such solutions 
also sit at the black hole threshold, but they generically
have two or more growing modes, so will {\em not} be  
intermediate attractors for {\em generic} 1-parameter interpolating
families.  However, there is evidence~\cite{LIE98} that at 
least some of these solutions may be constructed via carefully 
constructed families of initial data.

Gundlach's perturbative analysis of the EMKG system~\cite{GUN_EMKG}
also lead him to predict the existence of a universal ``wiggle'' in 
the mass-scaling law~(\ref{MASSSCALE}).  Hod and Piran presented 
similar arguments and numerical evidence for the oscillation
in~\cite{HOD_FS}.  It is probable that this wiggle is visible 
in Figure~\ref{FIG4}, but the systematic errors in the finite-difference 
computations which produced those results preclude a definitive
statement.

Finally in interesting recent work, Garfinkle~\cite{GAR97} considered
the separation of Einstein's equations into pieces governing
(1) the dynamics of the overall scale of a self-gravitating
system, and (2) the dynamics of the ``scale invariant'' part 
of the metric.  As part of his study he constructed a model
non-linear dynamical system (with three continuous degrees
of freedom) which exhibited many of the features seen in 
Type II critical collapse.

\begin{figure}[t]
\centerline{\epsfxsize=100mm\epsfbox{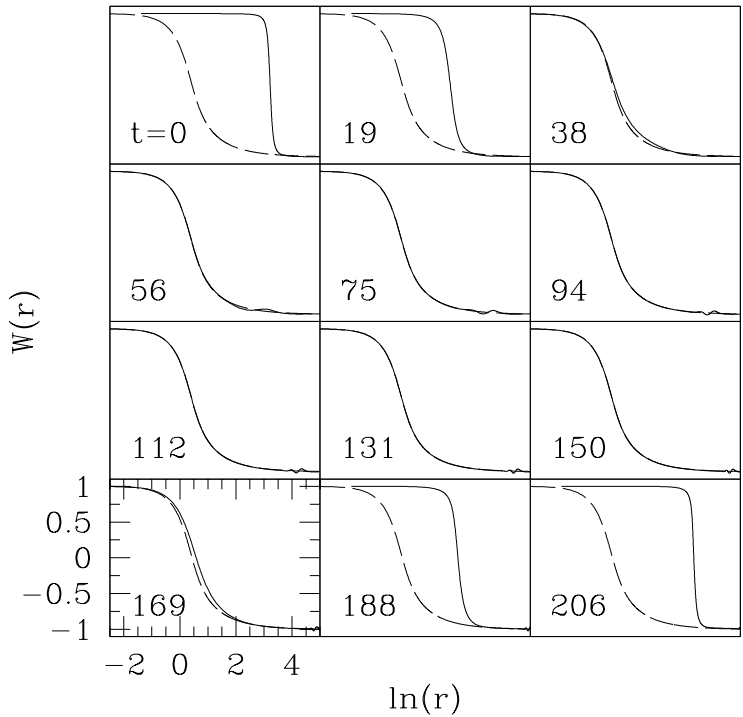}}
\caption{Evolution of the Yang-Mills potential, $W(r)$, in 
a near-critical evolution at the Type I transition in 
Einstein-Yang-Mills collapse.  Solid lines show the 
dynamical evolution of $W(r)$, while the dashed lines 
superimpose the static, $n=1$, Bartnik-Mckinnon solution. 
The initial data is an incoming ``kink'' which implodes to the 
center (frames 1-3), sheds off Yang-Mills radiation and asymptotes 
to the static solution (frames 4-9).  Because this evolution 
is sub-critical, the static configuration then completely 
disperses (frames 10-12) leaving flat spacetime behind.  In
a marginally super-critical case, a black hole containing essentially 
all of the mass of the static solution will form.
}
\label{FIG6}
\end{figure}

\section{Type I Behaviour: Einstein-Yang-Mills Collapse\label{SEC_EYM}}
In contrast to the recently discovered Type II transitions, 
Type I black-hole transitions are, of course, very 
familiar to astrophysicists in the context of the 
instability which generically arises in sequences of 
static stellar models parametrized, for example, by 
central density.  A somewhat different Type I transition
has recently been observed in the spherically symmetric 
collapse of an SU(2) Yang-Mills field~\cite{CHO96a}.
The equations of motion for this system are very similar 
to those for the EMKG model, with a single function, $W(r,t)$
(the ``Yang-Mills potential'') playing the role of the 
scalar field.  However, largely due to the non-trivial 
vacuum structure of the theory, the phenomenology here is richer 
than in the scalar case.  In particular, Bartnik
and Mckinnon~\cite{BAR88} showed that the Einstein-Yang-Mills
(EYM) model admits a countable infinity of regular, static 
solutions, conveniently labeled by the number, $n$, of 
zeros of the static profile
$W_n(r)$ ($W$ must be $\pm 1$ at the origin and at infinity).

All of the Bartnik-Mckinnnon solutions, which may be viewed as
resulting from a balance between the attractive 
gravitational interaction and the repulsive YM self-interaction,
were soon demonstrated to be unstable~\cite{STR90}, and, in fact, within
the specific {\em ansatz} in which they were originally constructed, 
were shown to have precisely $n$ unstable modes.  This suggested to Bizo\'n that
the $n=1$ solution could be constructed as a Type I critical solution 
from the dynamical evolution of suitably constructed one-parameter families
of initial data.  This indeed turned out to be the case~\cite{CHO96a},
and Figure~\ref{FIG6} shows the evolution of a configuration near the 
Type I transition.  As with the Type II solutions described
above, the critical solution here has exactly one unstable mode in
perturbation theory---tuning of initial data controls the effective
level of that mode in the initial data, and as one approaches 
the critical point, the unstable $n=1$ Bartnik-Mckinnon solution 
persists for a proper time 
\begin{equation}
	T \sim -\sigma \ln \left\vert p - p^\star \right\vert,
\end{equation}
where in complete analogy to the Type II analysis, the scaling 
exponent, $\sigma$, is simply the reciprocal of the real part of the 
eigenvalue corresponding to the unstable mode. 

Interestingly, Type II behaviour is also seen in the EYM model~\cite{CHO96a}
with $\Delta = 0.74$ and $\gamma=0.20$, providing rather definitive evidence 
of the non-universality of mass-scaling exponents.  Furthermore, considering
suitable two-parameter families of solutions, ${\cal S}[p_1,p_2]$, one 
generically finds a co-existence point of the two types of threshold 
behaviour in the $(p_1,p_2)$ plane.  Phenomena analogous to all
those described in this section have also been recently reported 
by Brady, Chambers and Goncalves~\cite{BRA97} in the context of 
the collapse of a {\em massive} scalar field.

\begin{figure}[t]
\begin{center}
\vspace{-10mm}
\epsfxsize=120mm
\epsfbox{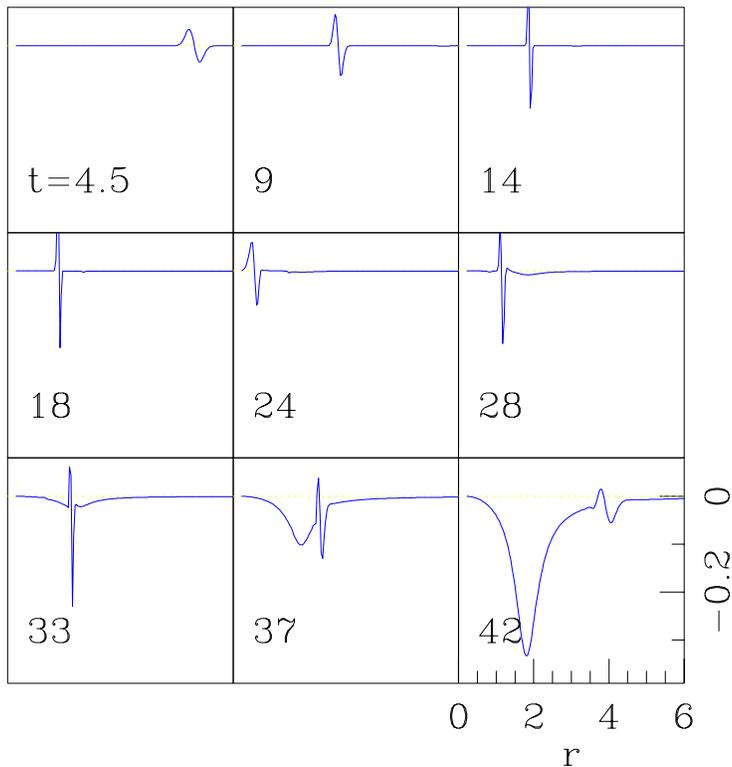}
\end{center}
\caption{Excitation of the growing mode of the static, irregular scalar 
field critical solution found by various authors~[40,41,42].
and recently discussed by van Putten~[43].  Here, the 
dynamical evolution of the quantity 
$r a \alpha^{-1}\partial \phi/\partial t$, which vanishes for the 
static solution, is plotted.  The initial data contains a
perturbation in the scalar field, which travels inward, bounces 
off the irregular origin between $t=18$ and $t=28$, then escapes
to infinity.  As it propagates through the region of the spacetime 
corresponding to the near-horizon region in a Schwarzschild spacetime,
the perturbing pulse excites the single, growing mode of the 
static solution.  Through judicious choice of perturbations, the 
growing mode can be excited with either sign---one sign leads 
to black hole formation, the other apparently leads to dispersal.
Also note the blue-shifting and red-shifting of 
the perturbation (indicative of the strong-field nature of 
the static solution) as it travels inwards and outwards respectively.
}
\label{FIG7}
\end{figure}
 
\section{An Irregular Critical Solution\label{SEC_IRREG}}
A final and novel example~\cite{CHO97} of black-hole critical behaviour is 
provided by a family of solutions (family parameter $\epsilon$) of the 
EMKG system which has
been independently discovered over the years by many 
authors~\cite{BUC59,BRA61,WIN68}, 
most recently by van Putten~\cite{VAN96}.  These solutions are 
spherically-symmetric and static, with an irregular origin, and have 
the feature that as the adjustable parameter $\epsilon$ is tuned to 0, 
the region of spacetime exterior to the effective support of the 
scalar field becomes a better and better approximation of the 
Schwarzschild geometry.
However, none of the solutions exhibit event horizons, and 
because of the difficulty in numerically treating horizon-containing 
spacetimes, van Putten proposed that the solutions might be useful
for ``mocking up'' black holes.  Unfortunately (from the point of 
view of that proposal), the solutions are unstable, and thus 
likely to be of little use as ``approximate black holes''.  
However, as with the 
other threshold solutions described above, these solutions
appear to be minimally unstable---that is, they seem to have 
a single unstable eigenmode in perturbation theory and, furthermore
(see Figure~\ref{FIG7}), appear to sit at the threshold of black hole
formation.

\section{Angular Momentum in Critical Collapse\label{SEC_ANG}}
Very recent work by Gundlach~\cite{GUN_AXI1,GUN_AXI2} addresses the 
important issue of the role of angular momentum at the threshold
of black hole formation.  In~\cite{GUN_AXI1},
non-spherical perturbations about the spherically-symmetric
Evans-Coleman radiation-fluid ($P=\rho/3$) critical solution are 
considered, with the conclusion that there are {\em no} growing
non-spherical modes.  This strongly suggests that, at least for small
departures from spherical symmetry, the Evans-Coleman solution 
will persist at the black hole threshold.  In~\cite{GUN_AXI2}
Gundlach investigates the dominant non-spherical mode ($\ell=1$),
which contributes to the angular momentum $L$ of the system
and predicts that the 
overall amplitude of the spin should scale as 
\begin{equation}
	L \propto \left\vert p - p^\star \right\vert^\mu \, ,
\end{equation}
with $\mu \approx 0.80$, but also that the spin vector will
rotate in space as a function of $p - p^\star$, with a 
universal frequency $\omega\approx 0.23$.  It will be extremely 
interesting to investigate these predictions, and related issues,
using full simulations in axisymmetry.

\section{Quantum Mechanical Effects\label{SEC_QUANTUM}}
The nature of Type II critical solutions---particularly the 
fact that they describe strong-field, gravitationally-mediated 
dynamics down to arbitrarily small scales---makes it natural 
to study the impact of matter-quantization on criticality. 
At the current time, the overall picture is murky, even 
given the restriction to semi-classical approaches. 
Here, there are at least two heuristic viewpoints.  On
the one hand, quantization will generally introduce a length
scale to the otherwise scale-free dynamics at a Type II 
threshold.  Thus, one might expect a mass gap to appear 
at threshold, and there is some evidence~\cite{CHI97,BOS96,PEL97}
for this.  On the other hand,  the predominant semi-classical 
effect which will be encountered is probably Hawking radiation,
which, one could argue, should merely ``renormalize'' the 
classical competition in the model, suggesting that the 
transition remains Type II~\cite{STR94,PIR97}.   
Despite the current confusion, one thing does seem clear---because 
Type II solutions are naturally strong-field on arbitrarily 
small scales, they {\em demand} to be quantized in a way
that the Schwarzschild solution, for example (where one must 
go in and, by hand, tune $M$ to zero to achieve unbounded 
curvatures) does not.  

\section{Conclusions\label{SEC_CONC}}
In a few short years, the study of the threshold of black hole 
formation has developed into an active and (at least to some 
of us!) exciting sub-field of relativity.  Although the 
initial results are now fairly well understood, it is 
clear that there remains an extremely rich phenomenology waiting 
to be explored, particularly in the context of 
non-spherically-symmetric collapse.  The reader may rest
assured that vigorous efforts to explore this new territory are 
currently underway.

\section{Acknowledgements}
It is a pleasure to thank my collaborators S.L.~Liebling, 
E.W.~Hirschmann, P.~Bizo\'n and T.~Chmaj.  The financial 
support of NSF PHY9722068 and a Texas Advanced Research
Project grant is also gratefully acknowledged.  Finally, I wish 
to thank the conference organizers for their substantial efforts in 
putting together such a wonderful meeting.

\end{document}